\documentclass[journal]{IEEEtran}

\usepackage{amsmath,amssymb,graphicx,hyperref}
\usepackage{microtype}
\usepackage{graphicx}
\usepackage{subfigure}
\usepackage{booktabs} 
\usepackage{array}

\usepackage{algorithm}
\usepackage{algpseudocode}
\usepackage[subtle]{savetrees}
\usepackage{makecell,xcolor}
\usepackage{hyperref}
\usepackage{multirow}
\usepackage[table]{xcolor}
\usepackage{adjustbox}
\usepackage{graphicx}    
\usepackage{anyfontsize} 
\usepackage{booktabs}    
\usepackage[table]{xcolor} 
\usepackage{makecell}    

\definecolor{top1}{rgb}{0.8, 0.5, 0.5}    
\definecolor{top2}{rgb}{0.95, 0.7, 0.7}   
\definecolor{top3}{rgb}{0.98, 0.9, 0.9}  %
\ifCLASSINFOpdf
\else
\fi
\usepackage{caption}
\begin{document}
%
\title{Unified Diffusion Refinement for Multi-Channel Speech Enhancement and Separation}
%
%
%

\author{Zhongweiyang~Xu,
        Ashutosh~Pandey,
        Juan~Azcarreta,
        Zhaoheng~Ni,
        Sanjeel~Parekh,
        Buye~Xu,
        and Romit~Roy~Choudhury,~\IEEEmembership{Fellow,~IEEE}
\thanks{Z.~Xu and R.~R.~Choudhury are with the Department of Electrical and Computer Engineering, University of Illinois at Urbana-Champaign, Champaign, IL 61820, USA (e-mail: zx21@illinois.edu, croy@illinois.edu).} 
\thanks{A.~Pandey, J. Azcarreta, Z. Ni, S. Parekh, and B. Xu are with the Reality Labs Research at Meta, Redmond, WA 98052, USA.}
}

\maketitle

\begin{abstract}
We propose Uni-ArrayDPS, a novel diffusion-based refinement framework for unified multi-channel speech enhancement and separation. Existing methods for multi-channel speech enhancement/separation are mostly discriminative and are highly effective at producing high-SNR outputs. However, they can still generate unnatural speech with non-linear distortions caused by the neural network and regression-based objectives. To address this issue, we propose Uni-ArrayDPS, which refines the outputs of any strong discriminative model using a speech diffusion prior. Uni-ArrayDPS is generative, array-agnostic, and training-free, and supports both enhancement and separation. Given a discriminative model's enhanced/separated speech, we use it, together with the noisy mixtures, to estimate the noise spatial covariance matrix (SCM). We then use this SCM to compute the likelihood required for diffusion posterior sampling of the clean speech source(s). Uni-ArrayDPS requires only a pre-trained clean-speech diffusion model as a prior and does not require additional training or fine-tuning, allowing it to generalize directly across tasks (enhancement/separation), microphone array geometries, and discriminative model backbones. Extensive experiments show that Uni-ArrayDPS consistently improves a wide range of discriminative models for both enhancement and separation tasks. We also report strong results on a real-world dataset. Audio demos are provided at \href{https://xzwy.github.io/Uni-ArrayDPS/}{https://xzwy.github.io/Uni-ArrayDPS/}.

\end{abstract}

\begin{IEEEkeywords}
Diffusion, Array Signal Processing, Multi-channel Speech Enhancement, Source Separation
\end{IEEEkeywords}

%
\IEEEpeerreviewmaketitle

\section{Introduction}
When multiple speakers talk simultaneously in a noisy room, the microphones record mixtures of the speakers' voices and environmental noise. This is known as the cocktail party problem~\cite{cocktail1, cocktail2}, where the goal is to extract clean speech sources from noisy mixtures. Speech enhancement typically assumes a single active speaker, whereas speech separation assumes multiple speakers speaking simultaneously. Deep learning--based supervised methods have shown remarkable potential for both speech enhancement~\cite{zheng2023sixty} and separation~\cite{araki202530+, wang_supervised_2018}. Most of these methods are discriminative and are trained end-to-end to directly map noisy mixture features to clean speech features. A regression loss is typically used as the training objective for enhancement, while speech separation further incorporates permutation-invariant training (PIT) to compute the loss for separated sources. Although these discriminative models achieve strong performance on objective metrics such as signal-to-noise ratio (SNR), they often introduce non-linear distortions due to neural network architectures, regression-based training objectives, and the ill-posed nature of speech enhancement and separation. These distortions not only degrade perceptual quality~\cite{distortion} but also reduce intelligibility~\cite{rethinkingdistortions}. This phenomenon is more pronounced in extremely noisy, low-SNR environments~\cite{spear, rethinkingdistortions, lavoce}.

In addition to discriminative methods, generative enhancement and separation approaches have shown strong potential for improving perceptual quality~\cite{sgmse, lutati2024separate}. For speech enhancement, \cite{cdiff, usee} condition a speech diffusion model on noisy speech. SGMSE~\cite{sgmse} starts the diffusion process from a mixture of noisy speech and Gaussian noise, and FlowSE~\cite{flowse} further extends this idea with flow matching. For speech separation, DiffSep~\cite{scheibler2023diffusion} tailors a stochastic differential equation (SDE) for source separation, and FLOSS~\cite{scheibler2025source} improves it with flow matching. Although these methods achieve strong perceptual quality, their objective metrics are often substantially worse than those of state-of-the-art (SOTA) discriminative methods for both enhancement and separation. Motivated by this gap, StoRM~\cite{storm} uses a discriminative enhancement model's output to initialize the diffusion model, and Diffiner~\cite{diffiner} uses a diffusion denoising restoration model (DDRM)~\cite{ddrm} to refine single-channel discriminative enhancement outputs. Similarly, for separation, combining discriminative and generative methods can yield the best performance~\cite{lutati2024separate}. However, these hybrid approaches have so far been limited to single-channel speech enhancement and separation.

Compared with the single-channel setting, multi-channel speech enhancement and separation can leverage spatial information, since speech and noise sources typically arrive from different directions. Spatial filtering (beamforming) enables effective separation of different sources~\cite{consolidated}. Similar to the single-channel case, discriminative models have also shown remarkable progress in multi-channel speech enhancement and separation. These architectures are designed to exploit spatial information either in the waveform domain~\cite{pcm, tadrn, fasnet} or in the short-time Fourier transform (STFT) domain~\cite{uses, uses2, spatialnet, tfgridnet, mcwobeamforming}. They can be adapted for enhancement and separation with slightly different training objectives, where separation still requires permutation-invariant training. By exploiting spatial information, these multi-channel models can achieve superior enhancement and separation performance compared with their single-channel counterparts. Moreover, because microphone arrays come in a variety of configurations, some models are designed to be array-agnostic~\cite{fasnet, tadrn, uses, uses2}, and once trained, can generalize across different array geometries.

Despite carefully designed architectures for spatial processing, these discriminative methods can still produce non-linear distortions~\cite{spear}, degrading perceptual quality and intelligibility, similar to single-channel discriminative methods. One way to mitigate these distortions is to use a deep learning model's output to estimate a traditional spatial filter, such as the minimum variance distortionless response (MVDR) beamformer~\cite{consolidated}. Applying the estimated beamformer to the multi-channel mixtures can help enforce distortionless speech in the output. However, such linear beamformers often leave more residual noise, necessitating additional post-processing~\cite{consolidated}.

To further mitigate distortions, there is a growing trend toward using generative models for multi-channel enhancement and separation~\cite{mcdiff, mcdiff2, mcdiff3, arraydps, arraydps-refine}. \cite{mcdiff, mcdiff2} use conditional diffusion for multi-channel enhancement, but with limited performance. \cite{mcdiff3} uses a diffusion module to refine a beamformer output, but the diffusion component does not explicitly incorporate multi-channel spatial information. More recently, ArrayDPS~\cite{arraydps} proposes a diffusion posterior sampling (DPS)~\cite{dps} framework for unsupervised, generative, and array-agnostic multi-channel speech separation. It uses a pre-trained speech diffusion model and estimates each source's room acoustic transfer functions (ATF) jointly with the posterior-sampling process. Despite its unsupervised nature, it achieves separation performance on par with SOTA discriminative methods. The framework has also been extended to other multi-channel inverse problems~\cite{usddps}. However, ArrayDPS assumes white noise and thus cannot be directly applied to speech enhancement. In contrast, ArrayDPS-Refine~\cite{arraydps-refine} is proposed to refine any discriminative multi-channel speech enhancement model using a pre-trained speech diffusion model. It first uses a discriminative model's output to estimate the noise spatial covariance matrix (SCM), and then uses the estimated SCM to compute the multi-channel mixture likelihood for diffusion posterior sampling. Although ArrayDPS-Refine can improve discriminative models in a training-free manner, it does not support speech separation.

In this paper, we build on our previous work on ArrayDPS-Refine, which targets multi-channel enhancement as described above. We propose Uni-ArrayDPS, a training-free, generative, and array-agnostic framework that can refine any state-of-the-art discriminative multi-channel speech enhancement or separation model. Similar to ArrayDPS and ArrayDPS-Refine, Uni-ArrayDPS requires only a pre-trained clean-speech diffusion model. It supports universal refinement across discriminative backbones, microphone array geometries, and tasks (enhancement and separation). As in ArrayDPS-Refine, it first uses the discriminative model's enhanced/separated outputs to estimate the noise SCM, which is then used during diffusion posterior sampling.

We extensively evaluate Uni-ArrayDPS for multi-channel enhancement and separation. Experiments show that Uni-ArrayDPS significantly improves perceptual quality, intelligibility, and automatic speech recognition (ASR) across a range of discriminative models for both tasks. We also present results on real-world multi-channel speech enhancement, demonstrating Uni-ArrayDPS's effectiveness in real-world scenarios.

We summarize our contributions as follows. Compared with ArrayDPS-Refine, we further extend the refinement approach to multi-channel speech separation, enabling more universal multi-channel speech refinement. We also improve performance by interpolating discriminative and generative outputs. In addition, we expand experiments to stronger SOTA discriminative models and a real-world recorded dataset. Finally, we provide more detailed ablations on likelihood-guidance parameters, diffusion sampling steps, and strategies for combining discriminative and generative outputs. Overall, we show that Uni-ArrayDPS can outperform SOTA discriminative models in \textbf{perceptual}, \textbf{intelligibility}, and \textbf{ASR} metrics for both multi-channel speech enhancement and separation.
%
%
%
%

\section{Background and Problem Formulation}\label{sec:background}
In a noisy, reverberant acoustic environment, a $C$-channel microphone array records mixtures of $K$ speakers talking simultaneously. Let $X^k(\ell,f)\in\mathbb{C}$ denote the $k^{\text{th}}$ anechoic clean speech source recorded at the reference microphone ($c=1$) in the short-time Fourier transform (STFT) domain, where $k\in[1,K]$ is the source index, $\ell\in[0,L-1]$ is the STFT frame index, and $f\in[0,F-1]$ is the STFT frequency index. Then, the $C$-channel noisy mixtures recorded by the microphones are modeled as a sum of reverberant speech sources and environmental noise:
\begin{equation}\label{eq:signal_model}
Y_c(\ell,f) = \sum_{k=1}^{K}H^k_{c}(\ell,f) *_\ell X^k(\ell,f) + N_c(\ell,f),~~~~c\in[1, C]
\end{equation}
where $Y_c(\ell,f)\in\mathbb{C}$ denotes the STFT-domain noisy mixture recorded by the $c^{\text{th}}$ microphone, $H^{k}_{c}(\ell,f)\in\mathbb{C}$ denotes the STFT-domain room acoustic transfer functions (ATFs) from the $k^{\text{th}}$ speech source to the $c^{\text{th}}$ microphone, and $N_c(\ell,f)\in\mathbb{C}$ denotes the environmental noise recorded at the $c^{\text{th}}$ microphone. Here, $*_\ell$ denotes convolution across STFT frames, and the room ATF $H^k_{c}\in\mathbb{C}^{N_H\times F}$ is a multi-frame filter with frame length $N_H$. For convenience, we let $Y(\ell,f) = [Y_1(\ell,f), Y_2(\ell,f), ..., Y_C(\ell,f)]\in\mathbb{C}^{C}$, and similarly for and $N(\ell,f), H^{k}(\ell,f)\in\mathbb{C}^{C}$.
Similarly, we let $X^{1:K}(\ell,f) = [X^1(\ell,f), X^2(\ell,f), ..., X^K(\ell,f)]\in\mathbb{C}^{K}$. Thus, Eq.~\ref{eq:signal_model} can be written in short as:
\begin{equation}\label{eq:signal_model_short}
Y(\ell,f) = \sum_{k=1}^{K}H^{k}(\ell,f) *_\ell X^k(\ell,f) + N(\ell,f)
\end{equation}

In the context of multi-channel speech enhancement, we assume $K=1$ and the goal is to extract $X^1$ given $Y$ (i.e., to sample from $p(X^1\mid Y)$). For multi-channel speech separation, the goal is to extract $X^{1:K}$ given $Y$ (i.e., to sample from $p(X^{1:K}\mid Y)$).

In multi-channel speech enhancement and separation, spatial-domain information is extremely crucial~\cite{spatialnet, consolidated}, so we make a spatially Gaussian assumption about the multi-channel noise $N$. We assume that $N(\ell, f)$ follows a zero-mean complex Gaussian distribution $ \mathcal{CN}\!\big(0, \Phi_{\text{NN}}(\ell,f)\big)$, where $\Phi_{\text{NN}}(\ell,f) = \mathbb{E}\big[\, N(\ell,f) N(\ell,f)^{H} \,\big]$ denotes the noise spatial covariance matrix (SCM). Given this noise assumption and Eq.~\ref{eq:signal_model_short}, we can write the likelihood of the noisy mixtures as:
\begin{align}
&p\big(Y(\ell,f)\,|\,H^{1:K}(\ell,f), X^{1:K}(\ell,f)\big)\nonumber\\
=  ~&\mathcal{CN}\!\Big(Y(\ell, f); \sum_{k=1}^K H^k(\ell,f) *_\ell X^k(\ell,f), \Phi_{\text{NN}}(\ell,f)\Big),\label{eq:likelihood}
\end{align}
where in Eq.~\ref{eq:likelihood}, $Y(\ell, f)$ follows complex Gaussian with the mean to be the multi-channel mixture of reverberant sources, and the covariance to be the noise spatial covariance.

\subsection{Diffusion Model}\label{sec:diffusion}
Diffusion models~\cite{ddpm, score, edm, improved_ddpm} have shown remarkable progress in generative modeling across multiple domains, including speech generation~\cite{diffwave}. A diffusion model first defines a forward diffusion process that gradually adds noise to clean data, and then generates samples by learning to remove Gaussian noise step by step.

We follow the Denoising Diffusion Probabilistic Model (DDPM)~\cite{ddpm, improved_ddpm} formulation. Starting from a data distribution $p_{\text{data}}(x_0)$, a forward diffusion process gradually transforms the clean signal $x_0$ to $x_1, x_2, ..., x_T$ as follows:
\begin{align}
x_t = \sqrt{\alpha_t} x_{t-1} + \sqrt{\beta_t} \epsilon_t,~~\epsilon_t\sim\mathcal{N}(0, I),~~t\in[1, T]\label{eq:diffusion_forward}
\end{align}
where $t$ is the diffusion time step, $\beta_t\in(0,1)$ is a pre-defined noise variance schedule to determine the amount of noise added in different diffusion steps. Then DDPM further defines $\alpha_t:=1-\beta_t$ which gradually scales $x_t$ at each diffusion step. From the forward process in Eq.~\ref{eq:diffusion_forward}, it is equivalent to directly transform $x_0$ to $x_t$ by 
\begin{equation}
x_t = \sqrt{\bar\alpha_t}\,x_0
      + \sqrt{1-\bar\alpha_t}\,\epsilon,\qquad \epsilon\sim\mathcal{N}(0,I).\label{eq:diffusion_forward2}
\end{equation}
where $\bar\alpha_t:=\prod_{s=1}^t \alpha_s$. As $t\rightarrow~T$, $\sqrt{\bar\alpha_t}\rightarrow~0$, so that finally $x_T$ almost becomes Gaussian noise with distribution $\mathcal{N}(0, I)$.

To generate a sample from $p(x_0)$, DDPM learns to reverse the forward diffusion process. Starting from a noise $x_T\sim\mathcal{N}(0, I)$, the sampling process reverses each diffusion step (from $t=T$ to $t=0$) by sampling from a learned posterior $p_\theta(x_{t-1}\!\mid x_t)$, until a clean sample $x_0$ is sampled. The learned reverse posterior is modeled as:
\begin{align}
p_\theta(x_{t-1}\!\mid x_t)&=\mathcal{N}\!\big(\mu_\theta(x_t,t),\,\sigma^2_t I\big),\label{eq:posterior}\\
\text{where~}\mu_\theta(x_t,t) &= \frac{1}{\sqrt{\alpha_t}}
   \!\left(
     x_t - \frac{\beta_t}{\sqrt{1-\bar\alpha_t}}\,
     \epsilon_\theta(x_t,t)
   \right),\label{eq:mu}\\
   \text{and~}\sigma_t&=\sqrt{\frac{1-\Bar{\alpha}_{t-1}}{1-\Bar{\alpha}_{t}}\beta_t}.\label{eq:sigma}
\end{align}
As shown in Eq.~\ref{eq:mu}, $\epsilon_\theta(x_t, t)$ is a neural network trained to estimate the noise $\epsilon$ in Eq.~\ref{eq:diffusion_forward2}. Thus, the training objective is to minimize:
\begin{align}
\mathbb{E}_{t,x_0\sim p_{\text{data}},\epsilon\sim\mathcal{N}(0,I)}\!
\Big[
\big\|
\epsilon - \epsilon_\theta(
\sqrt{\bar\alpha_t}\,x_0 + \sqrt{1-\bar\alpha_t}\,\epsilon,\; t)
\big\|_2^2
\Big].\label{eq:diffusion_loss}
\end{align}
Since the noise $\epsilon$ in Eq.~\ref{eq:diffusion_forward2} can be estimated by $\epsilon_\theta$ for any $x_t$, it also allows us to estimate $x_0$ from $x_t$ using the estimated noise, which can be shown to be a Minimum Mean Square Error (MMSE) denoiser:
\begin{equation}\label{eq:mmse}
    \mathbb{E}[x_0 \!\mid x_t] \simeq \hat{x}_0(x_t,t) = \frac{x_t - \sqrt{1-\bar\alpha_t}\,\epsilon_\theta(x_t,t)}{\sqrt{\bar\alpha_t}}
\end{equation}
Note that this MMSE estimator is one-step, which allows a direct estimation from $x_t$, and thus the denoised result would not be a realistic clean signal, but a smoothed and denoised signal.

Theoretically equivalent to DDPM, score-based diffusion~\cite{score, edm} formulates the forward diffusion process and reversal diffusion process as stochastic differential equations (SDE). When $T\rightarrow~\infty$, the diffusion forward process described by Eq.~\ref{eq:diffusion_forward} becomes the forward SDE below:
\begin{align}
    \mathrm{d}x_t &= -\frac{1}{2} \beta_t x_t \mathrm{d}t + \sqrt{\beta_t} \mathrm{d}w \label{eq:sde_forward} 
\end{align}
where $w$ in Eq.~\ref{eq:sde_forward} is the Wiener process. Similarly, the reversal diffusion process for sampling becomes another SDE:
\begin{align}
    \mathrm{d}x_t &= \left[ -\frac{1}{2} \beta(t) x_t - \beta(t) \nabla_{x_t} \log p_t(x_t) \right] \mathrm{d}t + \sqrt{\beta(t)} \mathrm{d}w \label{eq:sde_reverse}
\end{align}
In Eq.~\ref{eq:sde_reverse}, the score function $\nabla_{x_t} \log p_t(x_t)$ is usually approximated by a neural network $s_\theta(x_t, t)$, trained with a conditional score matching loss~\cite{score}. With the score function, we then start from a noise $x_T~\in\mathcal{N}(0, I)$ and solve the SDE in Eq.~\ref{eq:sde_reverse} to get $x_0\sim~p_{\text{data}}$, similar to DDPM sampling mentioned before. The score function and the MMSE denoiser discussed in Eq.~\ref{eq:mmse} can be directly connected by Tweedie's Formula:
\begin{align}
    \mathbb{E}[x_0 \!\mid x_t] = \frac{1}{\sqrt{\bar\alpha_t}} x_t + (1-\bar\alpha_t)\, \nabla_{x_t}\log~p_t(x_t).\label{eq:tweedie}
\end{align}
From Eq.~\ref{eq:mmse} and Eq.~\ref{eq:tweedie}, there is a direct relationship between the score function and the noise estimator, which allows DDPM to also access the score estimator:
\begin{equation}\label{eq:noise_score}
    \nabla_{x_t}\log \;p_t(x_t)\simeq~s_\theta(x_t, t)=-\frac{1}{\sqrt{1-\Bar{\alpha}_t}}\epsilon_\theta(x_t, t).
\end{equation}


\subsection{Diffusion Posterior Sampling and ArrayDPS}\label{sec:dps}
This section gives a background of diffusion posterior sampling (DPS)~\cite{dps}, which tries to solve inverse problems using a pre-trained diffusion prior. Assume $y=A(x)+n$, where $x$ is the clean signal to recover, $A(\cdot)$ is a known degradation operator, and $n$ is the white noise with variance $\sigma^2_y$. To recover the clean signal $x$ from the noisy measurement $y$, DPS samples from $p(x|y)$ using a pre-trained score diffusion model for the clean signal $x$.

As discussed in Sec.~\ref{sec:diffusion}, to sample from $p_\text{data}(x)$, diffusion models need to train a diffusion noise estimator $\epsilon_\theta(x_t, t)$ or a score model $s_\theta(x_t, t)$ to approximate the score $\nabla_{x_t}\log p_t(x_t)$, and knowing one directly infers the other. However, to sample from $p(x|y)$, the posterior score $\nabla_{x_t}\log p_t(x_t|y)$ is needed, so it is decomposed using Bayes' theorem:
\begin{equation}\label{eq:bayes}
    \nabla_{x_t}~\log~p(x_t|y) = \nabla_{x_t}~\log~p(x_t) + \nabla_{x_t}~\log~p(y|x_t).
\end{equation}
Note that the prior score $\nabla_{x_t}~\log~p(x_t)$ can be directly approximated by the pre-trained diffusion model $s_\theta(x_t, t)$, but the likelihood score $\nabla_{x_t}~\log~p(y|x_t)$ is still unknown. DPS then proposes to estimate the likelihood score by:
\begin{align}
    \nabla_{x_t}~\log~p(y|x_t) &\simeq~\nabla_{x_t}~\log~p(y|\hat{x}_0(x_t, t))\label{eq:likelihood_score1}\\
    &=\frac{1}{2\sigma^2_y}\nabla_{x_t}\|y-A(\hat{x}_0(x_t, t))\|_2^2\label{eq:likelihood_score2}\\
   \text{where}~\hat{x}_0(x_t, t) &= \frac{x_t - {\sqrt{1-\Bar{\alpha}_t}}\epsilon_\theta(x_t, t)}{\sqrt{\bar\alpha_t}}\label{eq:tweedie2}
\end{align}
Eq.~\ref{eq:tweedie2} uses the MMSE estimator $\hat{x}_0(x_t, t)$ (cf. Eq.~\ref{eq:mmse}). The estimate $\hat{x}_0$ can then be used to compute the likelihood in Eq.~\ref{eq:likelihood_score1} and Eq.~\ref{eq:likelihood_score2}, since $p(y\mid x)=\mathcal{N}(y; A(x), \sigma^2_y)$.

DPS shows remarkable result for image and audio inverse problems~\cite{dps, pigdm, undiff, cqt, eloi1}. However, it makes two assumptions that are unrealistic: 1) the degradation operator $A(\cdot)$ is known in advance, and 2) the noise $n$ has an analytical distribution like Gaussian or Laplace. There are a few methods proposed to solve the first problem of unknown operator $A$, where some solve for $A$ during DPS~\cite{buddy2, eloi2, eloi3, bwe}, some model $A$ using another diffusion model~\cite{blinddps}, and some train a latent surrogate for $A$~\cite{coguide}. In our problem formulation in Eq.~\ref{eq:signal_model_short}, the degradation operator $A(\cdot)$ in DPS is the ATF filter and sum operation, where all the ATFs ($H^{1:K}$) are unknown. Also, the noise $N$ in Eq.~\ref{eq:signal_model_short} is environmental noise, and thus the distribution is also unknown.

\subsection{ArrayDPS and FCP}\label{sec:arraydps}
As discussed in Sec.~\ref{sec:dps}, DPS cannot be directly applied to multi-channel speech separation, and one reason is that all the room ATFs ($H^{1:K}$) are unknown. ArrayDPS~\cite{arraydps} solves this by estimating the room ATFs ($H^{1:K}$) using Forward Convolutive Prediction (FCP)~\cite{fcp} at each DPS step for likelihood approximation.

ArrayDPS is designed for multi-channel speech separation, which is unsupervised, generative, and array-agnostic. It also uses a pre-trained clean speech diffusion model for diffusion posterior sampling. Its signal model follows Eq.~\ref{eq:signal_model_short}, but only considers white noise, i.e., $N\sim\mathcal{N}(0, \sigma^2_Y I)$. Following DPS, ArrayDPS's goal is to sample from $p(X^{1:K}|Y)$, which needs the posterior score $\nabla_{X^{1:K}}\log~p(X^{1:K}|Y)$. Thus, the posterior score is first decomposed using Bayes theorem:
\begin{align}
    &\nabla_{X_t^{1:K}}\log~p(X_t^{1:K}|Y)\nonumber\\
    = &\sum_{k=1}^{K}\nabla_{X_t^k}\log~p(X_t^k)  + \nabla_{X_t^{1:K}}\log~p(Y|X_t^{1:K})\label{eq:arraydps_bayes}
\end{align}
where $X^k_t$ denotes the $k^{\text{th}}$ reference-channel clean source at diffusion step $t$.  In Eq.~\ref{eq:arraydps_bayes}, each speech source's prior score $\nabla_{X^t_k}\log~p(X^t_k)$ can be approximated by a pre-trained diffusion noise denoiser $\epsilon_\theta(X^k_t, t)$, following Eq.~\ref{eq:noise_score}, and then the likelihood score $\nabla_{X_t^{1:K}}\log~p(Y|X_t^{1:K})$ is approximated by:
\begin{align}
\nabla_{X_t^{1:K}}\log~p(Y|X_t^{1:K})&\simeq~\nabla_{X_t^{1:K}}\log~p(Y|\hat{X}^{1:K}_0, \hat{H}^{1:K})\label{eq:arraydps_likelihood_score}\\
=\frac{1}{\sigma^2_Y}&\nabla_{X_t^{1:K}}\left\|Y-\sum_{k=1}^K \hat{H}^k *_l \hat{X}_0^k\right\|_2^2\label{eq:arraydps_likelihood_score2}\\
\text{where}~\hat{X}^k_0=&\frac{X^k_t - \sqrt{1-\bar\alpha_t}\, \epsilon_\theta(X^k_t, t)}{\sqrt{\bar\alpha_t}},\label{eq:tweedie3}\\
\text{and}~\hat{H}^k = &\text{FCP}(\hat{X}^k_0, Y),~k\in[1,K].\label{eq:arraydps_fcp}
\end{align}
Same as Eq.~\ref{eq:tweedie2}, Eq.~\ref{eq:tweedie3} first denoises each speech source $X^{k}_t$, and then Eq.~\ref{eq:arraydps_fcp} uses the denoised source $\hat{X}^{k}_0$ and the multi-channel mixtures $Y$ to estimate the room ATFs $H^k$ from source $k$ to $C$ microphones. The Forward Convolutive Prediction (FCP) algorithm in Eq.~\ref{eq:arraydps_fcp} is analytical and differentiable, which will be discussed later. Finally, Eq.~\ref{eq:arraydps_likelihood_score} and Eq.~\ref{eq:arraydps_likelihood_score2} uses the estimated clean sources and the ATFs to estimate the likelihood score, which can then be plugged into Eq.~\ref{eq:arraydps_bayes} for diffusion posterior sampling.

ArrayDPS's main contribution is to use FCP to estimate the unknown ATFs for likelihood approximation, as in Eq.~\ref{eq:arraydps_fcp}. FCP~\cite{fcp} is an STFT-domain filter estimation algorithm, which takes an input signal and a target signal. Intuitively, it finds the best filter such that after filtering the input with the filter, the filtered result matches the target signal the most. The problem formulation is shown as below:
\begin{align}
    &\hat{H}^k_c = \underset{{H}^k_c}{\arg\min} \sum_{l,f} \frac{1}{\hat{\lambda}_{l,f}} \left|Y_c({l,f}) - \sum_{j=0}^{N_H-1} H^k_c({j,f}) {X}^k({l-j,f})\right|^2,\label{eq:fcp_obj}\\
    &\text{where}~\hat{\lambda}_{l,f} = \frac{1}{C} \sum_{c=1}^C |Y_{c}({l, f})|^2+ \gamma\cdot \max\limits_{l,f}\frac{1}{C} \sum_{c=1}^C |Y_{c}({l,f})|^2\label{eq:fcp_lambda}
\end{align}
As in Eq.~\ref{eq:fcp_obj}, FCP fromulates the filter estimation as a weighted least square problem, whose solution is analytical. $N_H$ is the number of frames of the ATFs, and the inverse weight $\hat{\lambda}_{l,f}$ is defined in Eq.~\ref{eq:fcp_lambda} to prevent overfitting to high-energy STFT bins. $\gamma$ is a hyperparameter to tune the inverse weight $\hat{\lambda}_{l,f}$.

\begin{figure*}[t]
    \centering
    \hspace*{0.5cm}%
    \includegraphics[width=1.0\linewidth]{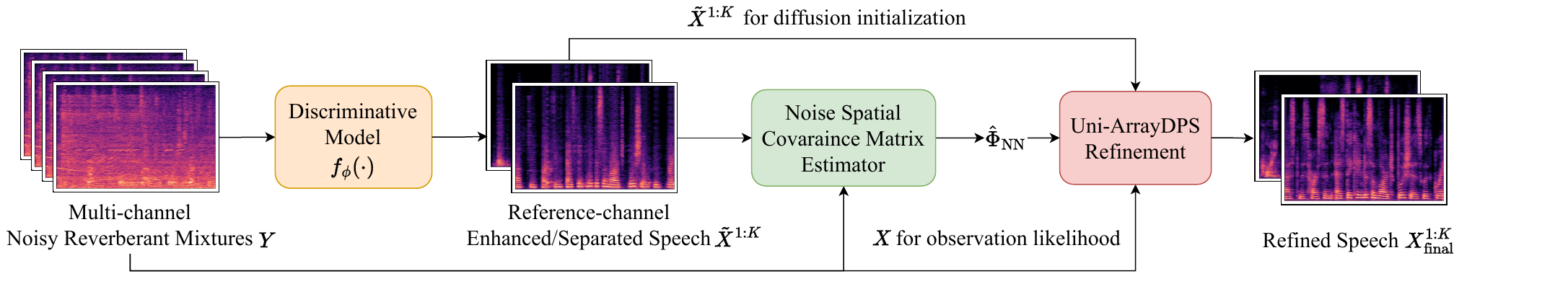}
    \caption{Uni-ArrayDPS Refinement Pipeline. }
    \label{fig:refinement_pipeline}
\end{figure*}

\section{Method}
As discussed in Sec.~\ref{sec:arraydps}, ArrayDPS supports diffusion posterior sampling for unsupervised multi-channel speech separation under the white noise assumption. It shows impressive separation results, even compared with many discriminative models. However, it does not work for real-world noisy environments, where noise sources coming from all directions can form a diffused sound field. Also, discriminative models like SpatialNet~\cite{spatialnet} show superior performance in multi-channel enhancement and separation. Thus, motivated by ArrayDPS and the effectiveness of discriminative models, we propose Uni-ArrayDPS, which uses an ArrayDPS-like module to further refine any discriminative multi-channel enhancement and separation models.

Uni-ArrayDPS's overall pipeline is shown in Fig.~\ref{fig:refinement_pipeline}. First, the discriminative enhancement or separation model $f_\phi(\cdot)$ processes the multi-channel mixtures $Y$ and outputs the estimated clean sources(s) $\tilde{X}^{1:K}=f_\phi(Y)$. Then, $Y$ and $\tilde{X}$ are used to estimate the noise spatial covariance matrix (SCM) $\hat{\Phi}_{\text{NN}}(\ell,f)\in\mathbb{C}^{C\times C}$. This estimated SCM allows the likelihood computation as mentioned in Sec.~\ref{sec:background} Eq.~\ref{eq:likelihood}. Finally, the Uni-ArrayDPS refinement module uses $\tilde{X}$ as an initialization, and uses $Y, \hat{\Phi}_{\text{NN}}(\ell,f)$ for arraydps-like diffusion posterior sampling.


\subsection{Spatial Covariance Matrix Estimation}\label{sec:scm}
After the discriminative enhancement/separation model estimates the reference-channel clean speech source(s) $\tilde{X}$, we first use it to estimate the multi-channel noise inside the noisy mixtures. Note that the noisy mixtures are multi-channel ($Y(\ell,f)\in\mathbb{C}^C$), while the denoised/separated source(s) are anechoic reference-channel ($\tilde{X}^k(\ell,f)\in\mathbb{C}$). Thus, we first use FCP to estimate the ATFs $\tilde{H}(\ell,f)\in\mathbb{C}^{C}$, and then get an estimate of the multi-channel reverberant clean source(s) $\tilde{X}^k_{\text{reverb}}(\ell,f)\in\mathbb{C}^C$:
\begin{align}
    \tilde{H}^k &= \text{FCP}(\tilde{X}^k, Y)\\
    \tilde{X}^k_{\text{reverb}} &= \tilde{H}^k *_\ell \tilde{X}^k~,k\in[1,K]
\end{align}
Then by subtracting the estimated multi-channel reverberant sources $\tilde{X}^k$ from the mixtures $Y$, we can get an estimate of the multi-channel noise $\tilde{N}(\ell,f)\in~\mathbb{C}^C$:
\begin{align}
    \tilde{N} = Y - \sum_{k=1}^K \tilde{X}_{\text{reverb}}^k
\end{align}
Using the estimated multi-channel noise, we then estimate the noise spatial covariance matrix in an exponential moving average manner:
\begin{align}
    \hat{\Phi}_{\text{NN}}(\ell,f) = \eta\hat{\Phi}_{\text{NN}}(\ell-1,f) + (1-\eta) \tilde{N}(\ell,f)\tilde{N}^H(\ell,f),\label{eq:ema}
\end{align}
where $\eta$ is a smoothing coefficient for the SCM update.


\subsection{Diffusion Model and Posterior Score Estimation}\label{sec:uniarraydps_derivation}
This section will first discuss Uni-ArrayDPS's diffusion model and then discuss how Uni-ArrayDPS updates at each diffusion step, using the estimated noise spatial covariance matrix (SCM) $\hat{\Phi}_{\text{NN}}$ mentioned in Sec.~\ref{sec:scm}.

Similar to ArrayDPS, Uni-ArrayDPS trains a DDPM-based prior diffusion for anechoic clean speech. However, instead of using waveform-domain~\cite{undiff, diffwave, arraydps} or STFT domain diffusion~\cite{diffiner}, we apply diffusion on the compressive STFT domain. Given the STFT of a clean signal, $X$, the compressive domain STFT is then $\bar{X}=|{X}|^{0.5}\,\exp\!\big(j\,\angle {X}\big)$. Since speech signals have much higher energy in low frequencies than high frequencies, the compression operation can effectively reduce the signal's dynamic range across frequencies. It has been shown that using these compressive STFT features can achieve better generation performance for diffusion models~\cite{edmsound}. Thus, the DDPM's forward diffusion process and the reversal sampling process are all in this compressive STFT domain.

Same as in Sec.~\ref{sec:diffusion}, the diffusion model trains a noise estimator $\epsilon_\theta(\bar{X}^k, t)$, which can then be used to sample from $p(\bar{X^k})$ using Eq.~\ref{eq:posterior} in Sec.~\ref{sec:diffusion}. Since Uni-ArrayDPS's goal is to sample from $p(\bar{X}^{1:K}|Y)$, we need to approximate the posterior score $\nabla~{\bar{X}}^{1:K}~\log~p(\bar{X}^{1:K}|Y)$, following ArrayDPS's framework as in Sec.~\ref{sec:arraydps}. Thus, same as DPS and ArrayDPS, $\nabla_{\bar{X}^{1:K}}~\log~p(\bar{X}^{1:K}|Y)$ is first decomposed into source prior scores and a likelihood score:
\begin{align}
    &\nabla_{\bar{X}_t^{1:K}}\log~p(\bar{X}_t^{1:K}|Y)\nonumber\\
    = &\sum_{k=1}^{K}\nabla_{\bar{X}_t^k}\log~p(\bar{X}_t^k)  + \nabla_{\bar{X}_t^{1:K}}\log~p(Y|\bar{X}_t^{1:K})\label{eq:uniarraydps_bayes_method}
\end{align}
Each source prior score $\nabla_{\bar{X}_k}\log~\log~p(\bar{X_k})$ can be directly approximated by $s_\theta(\bar{X}^k_t,t)=-\frac{1}{\sqrt{1-\Bar{\alpha}_t}}\epsilon_\theta(\bar{X}^k_t, t)$ (Eq.
~\ref{eq:noise_score}), and then Uni-ArrayDPS further approximates the likelihood score:
\begin{align}
&\nabla_{\bar{X}_t^{1:K}}\log~p(Y|\bar{X}_t^{1:K})\nonumber\\
\simeq~&\nabla_{\bar{X}_t^{1:K}}\log~p(Y|\hat{X}^{1:K}_0, \hat{H}^{1:K}, \hat{\Phi}_{\text{NN}})\label{eq:uniarraydps_likelihood_score1}\\
=&-\tfrac{1}{2}\nabla_{\bar{X}^{1:K}_t} \sum\limits_{\ell,f}
      \hat{N}(\ell,f)^{\mathrm{H}}\,
      \widehat{\Phi}_{\mathrm{NN}}^{-1}(\ell,f)\,
      \hat{N}(\ell,f)\label{eq:uniarraydps_likelihood_score2},\\
\text{where}~&\hat{\bar{X}}^k_0= \frac{1}{\sqrt{\bar{\alpha}_t}}(\bar{X}^k_t-\sqrt{1-\bar{\alpha}_t}{\epsilon_\theta(\bar{X}^k_t, t)})),\label{eq:tweedie4}\\
&\hat{X}^k_0 =|\hat{\bar{X}}^k_0|^2\,\exp\!\big(j\,\angle \hat{\bar{X}}^k_0\big),\label{eq:to_STFT}\\
&\hat{H}^k =\text{FCP}(\hat{X}^k_0, Y),~k\in[1,K],\label{eq:uniarraydps_fcp}\\
\text{and}~&\hat{N} = Y - \sum_{k=1}^K \hat{X}^k_0.\label{eq:noise_est}
\end{align}
This approximation is very similar to Eq.~\ref{eq:arraydps_likelihood_score} to Eq.~\ref{eq:arraydps_fcp}, except Uni-ArrayDPS further considers real-world spatial noise. Eq.~\ref{eq:tweedie4} firsts denoises the noisy speech features $\bar{X}^k_t$ to $\hat{\bar{X}}^k_0$, and then Eq.~\ref{eq:to_STFT} transforms $\hat{\bar{X}}^k_0$ to the STFT domain signal $\hat{X}^k_0$. Then, Eq.~\ref{eq:uniarraydps_fcp} and Eq.~\ref{eq:noise_est} uses $\hat{X}^k_0$ to estimate the RTFs $\hat{H}^k$ and the spatial noise $\hat{N}$, respectively. Finally, Eq.~\ref{eq:uniarraydps_likelihood_score1} uses the estimated sources, ATFs, and noise SCMs, to estimate the likelihood using Eq.~\ref{eq:likelihood}, with Eq.~\ref{eq:uniarraydps_likelihood_score2} as the derivation result.

Using the likelihood score approximation in Eq.~\ref{eq:uniarraydps_likelihood_score2} and Eq.~\ref{eq:uniarraydps_bayes}, we can then approximate the posterior score as:
\begin{align}
    &\nabla_{\bar{X}_t^{1:K}}\log~p(\bar{X}_t^{1:K}|Y)\nonumber\\
    \simeq &\sum_{k=1}^{K}s_\theta(\bar{X}^k_t) -\tfrac{1}{2}\nabla_{\bar{X}^{1:K}_t} \left[\sum\limits_{\ell,f}
      \hat{N}(\ell,f)^{\mathrm{H}}\,
      \widehat{\Phi}_{\mathrm{NN}}^{-1}(\ell,f)\,
      \hat{N}(\ell,f)\right].\label{eq:uniarraydps_bayes}
\end{align}
We can then get the estimate the diffusion noise conditioned on $Y$: $\epsilon_\theta(\bar{X}^{k}_t, t | Y)$. Then, from the relationship between the noise estimator and the score (Eq.~\ref{eq:noise_score}), we can then derive the conditional noise estimator $\epsilon_\theta(\bar{X}^{k}_t, t | Y)$ as:
\begin{align}
    \epsilon_\theta(\bar{X}^{k}_t, t)+\tfrac{1}{2}\sqrt{1-\bar{\alpha}_t}\nabla_{\bar{X}^{1:K}_t} \left[\sum\limits_{\ell,f}
      \hat{N}(\ell,f)^{\mathrm{H}}\,
      \widehat{\Phi}_{\mathrm{NN}}^{-1}(\ell,f)\,
      \hat{N}(\ell,f)\right].\label{eq:condition}
\end{align}
Finally, for the DDPM reversal sampling from $p(\bar{X}^{1:K})$, we can use Eq.~\ref{eq:posterior} to Eq.~\ref{eq:sigma} with $\epsilon_\theta(\bar{X}^{k}_t, t | Y)$ to get the Uni-ArrayDPS's diffusion update:
\begin{align}
    \bar{X}^k_{t-1}&= \frac{1}{\sqrt{\alpha_t}}(\bar{X}^k_t-\frac{1-\alpha_t}{\sqrt{1-\bar{\alpha}_t}}{\hat{\epsilon}})+\frac{1-\alpha_t}{\sqrt{\alpha_t}}G+\sigma^2_tZ,\label{eq:update}\\
    \text{where}~G&=-\tfrac{1}{2}\nabla_{\bar{X}^{1:K}_t} \!\left[
      \sum\limits_{\ell,f}
      \hat{N}(\ell,f)^{\mathrm{H}}\,
      \widehat{\Phi}_{\mathrm{NN}}^{-1}(\ell,f)\,
      \hat{N}(\ell,f)
    \right],\label{eq:G}\\
    \text{and}~Z&\sim~\mathcal{CN}(0, 2I).\label{eq:complex_gaussian}
\end{align}
Eq.~\ref{eq:complex_gaussian} samples from complex Gaussian with variance $2$ because our diffusion is defined on the real and imaginary components of the compressive STFT features. Eq.~\ref{eq:update} is then the Uni-ArrayDPS's one step update for diffusion posterior sampling.

\begin{algorithm}[t]
\caption{Uni-ArrayDPS}
\label{alg:inference}
\begin{algorithmic}[1]
\Require $Y, \hat{\Phi}_{\text{NN}}, \tilde{X}^{1:K}, T', \xi, \alpha, \epsilon_\theta(\cdot), \{\sigma^2_t\}_{t=1}^{T'}, \{\alpha_t\}_{t=1}^{T'}, \{\bar{\alpha}_t\}_{t=1}^{T'}$
\State $\bar{X}^{1:K} \gets |\tilde{X}^{1:K}|^{0.5}\,\exp\!\big(j\,\angle \tilde{X}^{1:K}\big)$ {\footnotesize\Comment{\textcolor{magenta}{to compressive domain}}}
\State Sample $\epsilon\sim\mathcal{CN}(0,2I)$ {\footnotesize\Comment{\textcolor{magenta}{$\mathcal{N}(0, I)$ for both real and imag}}}
\State $\bar{X}^{1:K}_{T'}\gets \sqrt{\bar\alpha_{T'}}\,\bar{X}^{1:K}
      + \sqrt{1-\bar\alpha_{T'}}\,\epsilon$ {\footnotesize\Comment{\textcolor{magenta}{init. to step T'}}}
\For{$t = T', \dots, 1$}
    \For{$k = 1, \cdots, K$}
    \State $\hat{\epsilon} \gets \epsilon_\theta(\bar{X}^k_t, t)$ {\footnotesize\Comment{\textcolor{magenta}{diffusion model estimate noise}}}
    \State Sample $Z\sim\mathcal{CN}(0,2I)$ {\footnotesize\Comment{\textcolor{magenta}{$\mathcal{N}(0, I)$ for both real and imag}}}
    \State $\bar{X}^k_{t-1}\gets \frac{1}{\sqrt{\alpha_t}}(\bar{X}^k_t-\frac{1-\alpha_t}{\sqrt{1-\bar{\alpha}_t}}{\hat{\epsilon}})+\sigma^2_tZ$ {\footnotesize\Comment{\textcolor{magenta}{prior step}}}
    \State $\hat{\bar{X}}^k_0\gets \frac{1}{\sqrt{\bar{\alpha}_t}}(\bar{X}_t-\sqrt{1-\bar{\alpha}_t}\hat{\epsilon})$ {\footnotesize\Comment{\textcolor{magenta}{one-step denoising}}}
    \State $\hat{X}^k_0 \gets |\hat{\bar{X}}^k_0|^2\,\exp\!\big(j\,\angle \hat{\bar{X}}^k_0\big)$ {\footnotesize\Comment{\textcolor{magenta}{transform to STFT domain}}}
    \State $\hat{H}^k\gets\text{FCP}(\hat{X}^k_0, Y)$ {\footnotesize\Comment{\textcolor{magenta}{room ATF estimation}}}
    \State $\hat{X}^k_\text{reverb}\gets \hat{H}^k*_\ell \hat{X}^k_0$ {\footnotesize\Comment{\textcolor{magenta}{estimate multi-channel reverb. speech}}}
    \EndFor
    \State $\hat{N} = Y-\sum_{k=1}^K\hat{X}^k_{\text{reverb}}${\footnotesize\Comment{\textcolor{magenta}{estimate multi-channel noise}}}
{    \State $G\gets\nabla_{\bar{X}_t} \!\left[
      -\tfrac{1}{2}\sum\limits_{\ell,f}
      \hat{N}(\ell,f)^{\mathrm{H}}\,
      \widehat{\Phi}_{\mathrm{NN}}^{-1}(\ell,f)\,
      \hat{N}(\ell,f)
    \right]${\footnotesize\Comment{\textcolor{magenta}{likelihood score}}}}
\State $\displaystyle
\bar{X}_{t-1} \gets \bar{X}^{1:K}_{t-1} + \xi\,\frac{1-\alpha_t}{\sqrt{\alpha_t}}\,
G$ {\footnotesize\Comment{\textcolor{magenta}{likelihood step}}}
\EndFor
\State ${X}^{1:K}_0 \gets |\bar{X}^{1:K}_0|^2\,\exp\!\big(j\,\angle \bar{X}^{1:K}_0\big)$ {\footnotesize\Comment{\textcolor{magenta}{transform to STFT domain}}}
\For{$k = 1, \cdots, K$}
    \State $H^k_{\text{align}}\gets \text{FCP}(X^k_0, \tilde{X}^k)$ {\footnotesize\Comment{\textcolor{magenta}{single-frame filter est. for alignment}}}
    \State $X^k_{\text{align}}\gets H^k_{\text{align}}*_\ell X^k_0$
\EndFor
\State $X^{1:K}_{\text{final}} = \alpha \tilde{X}^{1:K} + (1-\alpha) X^{1:K}_{\text{align}}$ {\footnotesize\Comment{\textcolor{magenta}{interpolate}}}
\State \Return $X^{1:K}_{\text{final}}$ {\footnotesize\Comment{\textcolor{magenta}{return aligned signal}}}
\end{algorithmic}
\end{algorithm}
\subsection{Uni-ArrayDPS Algorithm}\label{sec:uniarraydps_algorithm}
This section then explains the Uni-ArrayDPS algorithm in detail, which is shown in Algorithm~\ref{alg:inference}. Uni-ArrayDPS takes the noisy multi-channel speech $Y$, the discriminative enhanced/separated speech source(s) $\tilde{X}^{1:K}$, the estimated noise SCM $\hat{\Phi}_{\text{NN}}$, and a few hyper-parameters as inputs. In line 1, $\tilde{X}^{1:K}$ is first transformed to the compressive STFT domain sources $\bar{X}^{1:K}$. Then, line 2-3 applies a forward diffusion process to $\bar{X}^{1:K}$, getting $\bar{X}^{1:K}_{T'}$, which is at diffusion step $T'\in[1, T]$. This is the same as ArrayDPS~\cite{arraydps}, where the DPS does not start from diffusion step ($t=T$), but from an intermediate diffusion step ($t=T'$) initialized from $\bar{X}^{1:K}$. Similar to Eq.~\ref{eq:complex_gaussian}, line 2 is also sampling from $\mathcal{CN}(0, 2I)$ because the diffusion takes place in the real and imaginary components of the signal.

Line 4-17 in Algorithm~\ref{alg:inference} shows the diffusion posterior sampling, which gradually transforms the initialization $\bar{X}^{1:K}_{T'}$ to the refined compressive STFT $\bar{X}^{1:K}_{0}$. The DPS update follows the derivation in Sec.~\ref{sec:uniarraydps_derivation}, where Eq.~\ref{eq:update} is the final update rule. For the specific implementation of Eq.~\ref{eq:update}, line 6 first estimates the diffusion noise, and then lines 7-8 apply a prior diffusion sampling step. Line 9 further denoises $\bar{X}^k_t$ using the estimated diffusion noise, to get the clean source estimate $\hat{\bar{X}}^k_0$. Then, line 10 transforms $\hat{\bar{X}}^k_0$ back to STFT domain, line 11 uses FCP to estimate the $k^{\text{th}}$ source's ATFs $\hat{H}^k$, and line 12 applies the ATFs to $\hat{X}^k_0$ to get an estimate of the multi-channel reverberant source $\hat{X}_{\text{reverb}}^k$. While all $K$ sources are processed, line 15 calculates the likelihood score as in Eq.~\ref{eq:G}. Finally, line 16 applies the likelihood score update, which is the same as in Eq.~\ref{eq:update}, except we add a $\xi>0$ hyper-parameter to control likelihood guidance. Although $\xi$ is empirical, it allows control of a trade-off between naturalness and hallucination, which we will later show in Sec.~\ref{sec:ablations}.

The DPS sampling output $\bar{X}^{1:K}_0$ is still in the compressive domain, so line 18 transforms it back to the STFT domain signal ${X}^{1:K}_0$. However, during sampling, the likelihood guidance does not constrain that ${X}^{1:K}_0$ would align with the reference-channel clean signal, which is a known problem in UNSSOR~\cite{unssor} and ArrayDPS~\cite{arraydps}. Thus, line 20 estimates a one-frame ($N_H=1$) filter that can align ${X}^{k}_0$ to the discriminative enhancement/separation output ${X}^{k}_0$. Since the discriminative model is trained to align with the reference-channel signal, ${X}^{k}_{\text{align}}$ will also be aligned. Finally, line 23 interpolates the final aligned outputs ${X}^{1:K}_{\text{align}}$ with the discriminative model's output $\tilde{X}^{1:K}$, output the final sources $X^{1:K}_{\text{final}}$. We call $\alpha$ in line 23 the discriminative-generative interpolation coefficient. This trick is also shown to be effective in other discriminative-generative hybrid approaches~\cite{hirano2023diffusion}.

\section{Experimental Setup}
This section discusses the Uni-ArrayDPS hyper-parameter configurations, all the discriminative model baselines, simulated and real-world datasets, and evaluation metrics.
\subsection{Uni-ArrayDPS Configurations}\label{sec:config}
For the prior diffusion model, we follow DDPM~\cite{ddpm} and use a noise schedule $\beta_t$ that increases linearly from $\beta_1=10^{-4}$ to $\beta_T=0.02$, with $T=1000$ steps. We use a 2-D U-Net with residual blocks as the architecture of our diffusion noise estimator $\epsilon_\theta(\bar{X}_t, t)$. The U-Net architecture is the same as the one used in Diffiner~\cite{diffiner}, modified from~\cite{improved_ddpm} to accommodate 2-D STFT. For the STFT in diffusion, we use an FFT size of 512, a hop size of 128, and a square-root Hann window. We pad the number of frames to 512 and remove the DC component, so the input to our U-Net is the real and imaginary channels of the noisy compressive STFT $\bar{X}_t$, with two channels, 256 frequency bins, and 512 frames. We train the diffusion model on about 220 hours of clean speech from the first DNS-Challenge~\cite{dns}. Each training sample is a 4-second, 16-kHz clean speech utterance, and we normalize each sample's waveform to $[-1,1]$. We use the Adam optimizer~\cite{adam}, with a learning rate of $10^{-4}$ and a batch size of 64, and train the model for $2.5\times 10^6$ steps on 8 H100 GPUs. We also use an exponential moving average (EMA) of the model weights with a decay of $0.9999$.

We estimate the noise SCM via the exponential moving average in Eq.~\ref{eq:ema}, using $\eta=0.95$. For Uni-ArrayDPS (see Algorithm~\ref{alg:inference} and Sec.~\ref{sec:uniarraydps_algorithm}), diffusion sampling begins at an intermediate step $T'$. We set this $T'$ to be $300$ by careful tuning, and study the effects of $T'$ by sweeping $T'\in\{100, 200, 300, 400, 500\}$. We also sweep the likelihood-guidance parameter $\xi$ to study the balance between prior-driven quality and likelihood-driven mixture fidelity, and evaluate $\xi\in\{0.4, 0.6, 0.8, 1.0, 1.2\}$. In Algorithm~\ref{alg:inference} line 11, for FCP's parameter in Eq.~\ref{eq:fcp_lambda}, we use a $N_H=13$-frame filter with $\gamma=10^{-3}$, matching the setting in ArrayDPS~\cite{arraydps}. For the alignment FCP (Algorithm~\ref{alg:inference} line 20), we set $N_H=1$ for single-frame alignment. For the discriminative-generative interpolation $\alpha$ in Algorithm~\ref{alg:inference} line 23, we find that $\alpha=0.5$ is a good default value, and we sweep through $\alpha\in\{0, 0.1, 0.3, 0.5, 0.7, 0.9, 1.0\}$ for ablations in the result section.

\subsection{Discriminative Baselines and Datasets}\label{sec:baseline}
For the discriminative baseline models, we use three array-agnostic models: FaSNet-TAC~\cite{fasnet}, TADRN~\cite{tadrn}, and USES2~\cite{uses2}, which once trained, can directly be applied to any microphone-array geometry. We also use one strong array-specific baseline model SpatialNet~\cite{spatialnet}, which can only work for a fixed mumber of microphones. Note all these models can be trained to support either enhancement or separation by changing model's number of output channels.

To support array-agnostic enhancement/separation, FaSNet-TAC employs transform-average-concatenate (TAC) to multi-channel time-domain signals. We use FaSNet-TAC's official implementation and configuration\footnote{https://github.com/yluo42/TAC/blob/master/FaSNet.py}. TADRN is a strong time-domain enhancement/separation model, which uses a triple-path attention architecture to process information across frames, chunks, and channels; we use the same MIMO configuration as in the original paper~\cite{tadrn}. USES2~\cite{uses2} is an STFT-domain array-agnostic competitive model, and we use the USES2-Comp configuration as in the original paper, following the official implementation\footnote{https://github.com/espnet/espnet}, with FFT size 512, STFT hop size 256, and a square-root Hann window. SpatialNet is a state-of-the-art STFT-domain model, which uses narrow-band channel-wise attention to fully exploit the spatial information. We use the SpatialNet-Large configuration in the original paper, following the official implementation\footnote{https://github.com/Audio-WestlakeU/NBSS}.

For discriminative model training, we create ad-hoc microphone array datasets for both multi-channel speech enhancement and separation. Both tasks have the exact same dataset simulation settings, except that enhancement simulates one target speaker and separation simulates two. To simulate a data sample, a shoe-box room is randomly drawn, with three dimensions uniformly sampled from $3\times3\times2$ to $10\times10\times5$ m. Similarly, we also uniformly sample the absorption coefficient from $0.3$ to $0.7$, resulting in a $T60\in[0.13, 0.55]$ s. We then sample the microphone-array position in the room randomly, and then the positions of 8 microphones are randomly sampled inside a sphere centered at the array position, with a radius of $0.1$ m. Thus, each sample's microphone array geometry is different in the dataset. We also randomly sample $8-16$ interference speakers to simulate bubble noise, and  $1-50$ noise sources to simulate diffused noise field. We sample $1$ target speaker source for the enhancement datasets and sample $2$ target speaker sources for the separation datasets. All sources' locations and the microphone center location is randomly sampled in the room. The speech and noise sources are all sampled from the DNS-Challenge dataset~\cite{dns}. We uniformly sample the signal-to-noise ratio to be from $[-10, 5]$ dB, and the signal-to-interference ratio to be from $[5, 10]$ dB (signal-to-interference ratio is defined to be the target speakers' energy over the interference speakers' energy). The acoustic simulation uses the image-source method~\cite{imagesource} (order 6) from the Pyroomacoustics toolbox~\cite{pyroomacoustics}. For both enhancement and separation datasets, we simulate $80,000$ $10$-second training samples, $1,000$ $4$-second validation samples, and $1,000$ $4$-second test samples.

For array-agnostic models including FaSNet-TAC, TADRN, and USES2, we train one model for enhancement and another for separation. During training, the number of channels in each batch is randomly sampled from $2$ to $8$, which allows training on variable number of channels. For SpatialNet model training, since one model cannot work for variable number of channels, we train 4 different models, for 4-channel enhancement, 4-channel separation, 8-channel enhancement, and 8-channel separation, respectively. During training, each model is only trained on a fixed number of channels. For all the models, we use the Phase Constrained Magnitude (PCM) loss~\cite{pcm_loss} as the training objective, which is a combination of time-domain loss and STFT magnitude loss. We use the anechoic clean speech as the training target. For separation training, we further use the permutation invariant training (PIT)~\cite{tadrn} to calculate the PCM loss. Adam optimizer with a learning rate of $10^{-4}$ is used. For FaSNet-TAC, TADRN, and SpatialNet, we use a batch size of 16 and train for 80 epochs. For USES2, we use a batch size of 8 and train for 40 epochs.

\section{Evaluation Results}\label{sec:result}
This section shows the evaluation results of all the discriminative baselines, and Uni-ArrayDPS's refinement over these baselines. We evaluate on both simulated and real-world datasets for multi-channel enhancement and separation.

\begin{table*}[t]
\scriptsize
\centering
\caption{Uni-ArrayDPS evaluation for multi-channel speech enhancement on the simulated adhoc microphone array dataset.}
\setlength\tabcolsep{0.7pt}
\renewcommand{\arraystretch}{0.7}
\label{tab:enhancement_simulated}
\begin{tabular}{@{}c c | c | c | ccccccc | ccccccc@{}}
\toprule
\multirowcell{2}{row} &
\multirowcell{2}{Methods} &
\multirowcell{2}{$\xi$} &
\multirowcell{2}{$\alpha$} &
\multicolumn{7}{c|}{\textbf{4-channel}} &
\multicolumn{7}{c}{\textbf{8-channel}} \\
\cmidrule(lr){5-11} \cmidrule(l){12-18}
& & & & STOI & eSTOI & { PESQ(NB/WB)} & SI-SDR & WER(\%) & DNSMOS & UTMOSv2 & STOI & eSTOI & {PESQ(NB/WB)} & SI-SDR & WER(\%) & {DNSMOS} & UTMOSv2 \\
\midrule
\rowcolor{gray!10}
A0 & Noisy & - & - & 0.622 & 0.347 & 1.38 / 1.07 & -6.7 & 78.9 & 1.47 & 1.85 & 0.622 & 0.347 & 1.38 / 1.07 & -6.7 & 78.9 & 1.47 & 1.85 \\
\midrule
A1 & TADRN \cite{tadrn} & - & - & 0.893 & 0.774 & 2.83 / 2.01 & 8.9 & 41.8 & 2.84 & 2.58 & 0.909 & 0.805 & 2.94 / 2.16 & 9.8 & 35.4 & 2.86 & 2.67 \\
\rowcolor{gray!10}
A2 & Refined TADRN & 0.4 & 0 & 0.907 & 0.803 & 2.95 / 2.17 & \textbf{10.0} & 36.7 & \textbf{2.91} & \textbf{2.97} & \textbf{0.925} & 0.835 & 3.09 / 2.36 & \textbf{11.0} & 29.0 & \textbf{2.91} & \textbf{3.02} \\
A3 & Refined TADRN & 0.4 & 0.5 & 0.906 & 0.801 & 2.98 / 2.20 & 9.8 & 35.2 & 2.90 & 2.83 & 0.922 & 0.830 & 3.10 / 2.38 & 10.7 & 29.3 & 2.90 & 2.90 \\
\rowcolor{gray!10}
A4 & Refined TADRN & 0.8 & 0.5 & 0.908 & 0.805 & \textbf{2.99} / 2.22 & 9.9 & 34.6 & 2.89 & 2.82 & 0.924 & 0.836 & 3.12 / 2.41 & 10.9 & 28.3 & 2.90 & 2.89 \\
A5 & Refined TADRN & 1.0 & 0.5 & \textbf{0.908} & \textbf{0.807} & 2.98 / \textbf{2.22} & 9.9 & \textbf{34.2} & 2.89 & 2.82 & \textbf{0.925} & \textbf{0.837} & \textbf{3.12} / \textbf{2.41} & 10.9 & \textbf{27.3} & 2.89 & 2.89 \\
\midrule
B1 & FaSNet-TAC \cite{fasnet} & - & - & 0.833 & 0.664 & 2.49 / 1.61 & 5.4 & 57.7 & 2.54 & 1.77 & 0.853 & 0.698 & 2.57 / 1.70 & 6.3 & 50.7 & 2.58 & 1.89 \\
\rowcolor{gray!10}
B2 & Refined FaSNet-TAC & 0.4 & 0 & \textbf{0.859} & \textbf{0.721} & \textbf{2.68} / \textbf{1.82} & \textbf{6.6} & 47.8 & \textbf{2.73} & \textbf{2.56} & \textbf{0.885} & \textbf{0.763} & \textbf{2.82} / \textbf{1.98} & \textbf{7.6} & 39.4 & \textbf{2.75} & \textbf{2.66} \\
B3 & Refined FaSNet-TAC & 0.4 & 0.5 & 0.854 & 0.706 & \textbf{2.68} / 1.79 & 6.2 & 48.0 & 2.67 & 2.21 & 0.875 & 0.743 & 2.79 / 1.93 & 7.2 & 41.0 & 2.70 & 2.35 \\
\rowcolor{gray!10}
B4 & Refined FaSNet-TAC & 0.8 & 0.5 & 0.857 & 0.712 & \textbf{2.68} / 1.80 & 6.3 & 47.5 & 2.65 & 2.18 & 0.878 & 0.748 & 2.79 / 1.94 & 7.2 & 39.7 & 2.69 & 2.28 \\
B5 & Refined FaSNet-TAC & 1.0 & 0.5 & 0.858 & 0.713 & \textbf{2.68} / 1.81 & 6.3 & \textbf{46.0} & 2.64 & 2.14 & 0.879 & 0.749 & 2.79 / 1.95 & 7.3 & \textbf{38.7} & 2.67 & 2.27 \\
\midrule
C1 & USES2 \cite{uses2} & - & - & 0.919 & 0.825 & 3.09 / 2.35 & 6.0 & 31.3 & 2.85 & 2.88 & 0.931 & 0.849 & 3.21 / 2.51 & 5.8 & 26.6 & 2.86 & 2.97 \\
\rowcolor{gray!10}
C2 & Refined USES2 & 0.4 & 0 & 0.918 & 0.826 & 3.07 / 2.35 & \textbf{6.6} & 30.3 & 2.88 & 3.03 & 0.934 & 0.854 & 3.21 / 2.54 & \textbf{6.4} & 24.0 & 2.89 & 3.09 \\
C3 & Refined USES2 & 0.4 & 0.5 & \textbf{0.926} & 0.839 & \textbf{3.18} / \textbf{2.50} & \textbf{6.6} & \textbf{26.9} & \textbf{2.90} & \textbf{3.06} & 0.938 & 0.862 & \textbf{3.30} / \textbf{2.67} & 6.3 & 23.0 & \textbf{2.91} & \textbf{3.12} \\
\rowcolor{gray!10}
C4 & Refined USES2 & 0.8 & 0.5 & \textbf{0.926} & \textbf{0.840} & 3.16 / 2.49 & \textbf{6.6} & 27.0 & 2.89 & 3.04 & 0.939 & 0.865 & \textbf{3.30} / \textbf{2.68} & 6.3 & \textbf{22.3} & 2.89 & 3.11 \\
C5 & Refined USES2 & 1.0 & 0.5 & \textbf{0.926} & \textbf{0.840} & 3.14 / 2.47 & \textbf{6.6} & 27.1 & 2.89 & 3.03 & \textbf{0.939} & \textbf{0.866} & 3.29 / 2.67 & 6.3 & 22.5 & 2.88 & 3.09 \\
\midrule
D1 & SpatialNet \cite{spatialnet} & - & - & 0.944 & 0.873 & 3.32 / 2.74 & 13.5 & 22.0 & 2.86 & 3.06 & 0.959 & 0.904 & 3.49 / 3.01 & 14.7 & 16.8 & 2.88 & 3.18 \\
\rowcolor{gray!10}
D2 & Refined SpatialNet & 0.4 & 0 & 0.938 & 0.863 & 3.24 / 2.62 & 13.3 & 22.8 & 2.90 & 3.12 & 0.954 & 0.894 & 3.41 / 2.88 & 14.4 & 17.1 & 2.90 & 3.16 \\
D3 & Refined SpatialNet & 0.4 & 0.5 & \textbf{0.946} & \textbf{0.879} & \textbf{3.38} / \textbf{2.84} & \textbf{14.0} & \textbf{20.7} & \textbf{2.91} & \textbf{3.17} & \textbf{0.961} & \textbf{0.909} & \textbf{3.55} / \textbf{3.12} & \textbf{15.2} & 15.6 & \textbf{2.92} & \textbf{3.25} \\
\rowcolor{gray!10}
D4 & Refined SpatialNet & 0.8 & 0.5 & \textbf{0.946} & \textbf{0.880} & 3.36 / 2.81 & \textbf{14.0} & \textbf{20.0} & 2.90 & 3.16 & \textbf{0.961} & \textbf{0.910} & 3.54 / 3.10 & \textbf{15.2} & \textbf{15.1} & 2.91 & 3.23 \\
D5 & Refined SpatialNet & 1.0 & 0.5 & \textbf{0.946} & 0.879 & 3.34 / 2.79 & \textbf{14.0} & \textbf{20.7} & 2.89 & 3.16 & \textbf{0.961} & 0.909 & 3.52 / 3.08 & \textbf{15.2} & 15.5 & 2.90 & 3.22 \\
\bottomrule
\end{tabular}
\end{table*}

\subsection{Evaluation Datasets and Metrics}\label{sec:datasets_metrics}
For evaluation, we use the simulated datasets discussed in Sec.~\ref{sec:baseline}, where we evaluate on both 4-channel (use first 4 channels) and 8-channel enhancement/separation. In addition to simulated dataset, we also evaluate 4-channel and 8-channel enhancement on the RealMan dataset~\cite{realman}. Following the official configuration~\footnote{https://github.com/Audio-WestlakeU/RealMAN}, we set the SNR range to be $-15$ to $5$ dB, speaker to be static, and the audio sample length to be 4-second. We also use the first 4 microphone or 8 microphone for enhancement evaluation.

We use extensive metrics to measure enhanced/separated speech's intelligibility, perceptual quality, and sample-level consistency. We use Short-Time Objective Intelligibility (STOI)~\cite{stoi}, extended STOI~\cite{estoi}, and word error rate (WER) or character error rate (CER) to measure speech intelligibilty. We use the Whisper~\cite{whisper} base model to get transcripts of both the enhanced/separated signal and the ground-truth clean signal, and then calculate the WER for the simulated datasets (in English), or the CER for the RealMan dataset (in Chinese). We further use Perceptual Evaluation of Speech Quality (PESQ)~\cite{pesq}, DNSMOS~\cite{dnsmos}, and UTMOSv2~\cite{utmosv2} to evaluate speech perceptual quality. For PESQ, we evaluate both the narrow-band (NB) and the wide-band (WB) metrics. Also, DNSMOS and UTMOSv2 are all non-intrusive metrics which does not need a reference clean signal. Lastly, we calculate SI-SDR~\cite{sisdr} to measure sample-level consistency.

\begin{figure*}[t]
    \centering
    \caption{Uni-ArrayDPS refinement performance with ablations on likelihood guidance $\xi$, diffusion starting step $T'$, and generative-discriminative interpolation coefficient $\alpha$. The result is shown on 4-channel speech enhancement.}
    \includegraphics[width=1.0\linewidth]{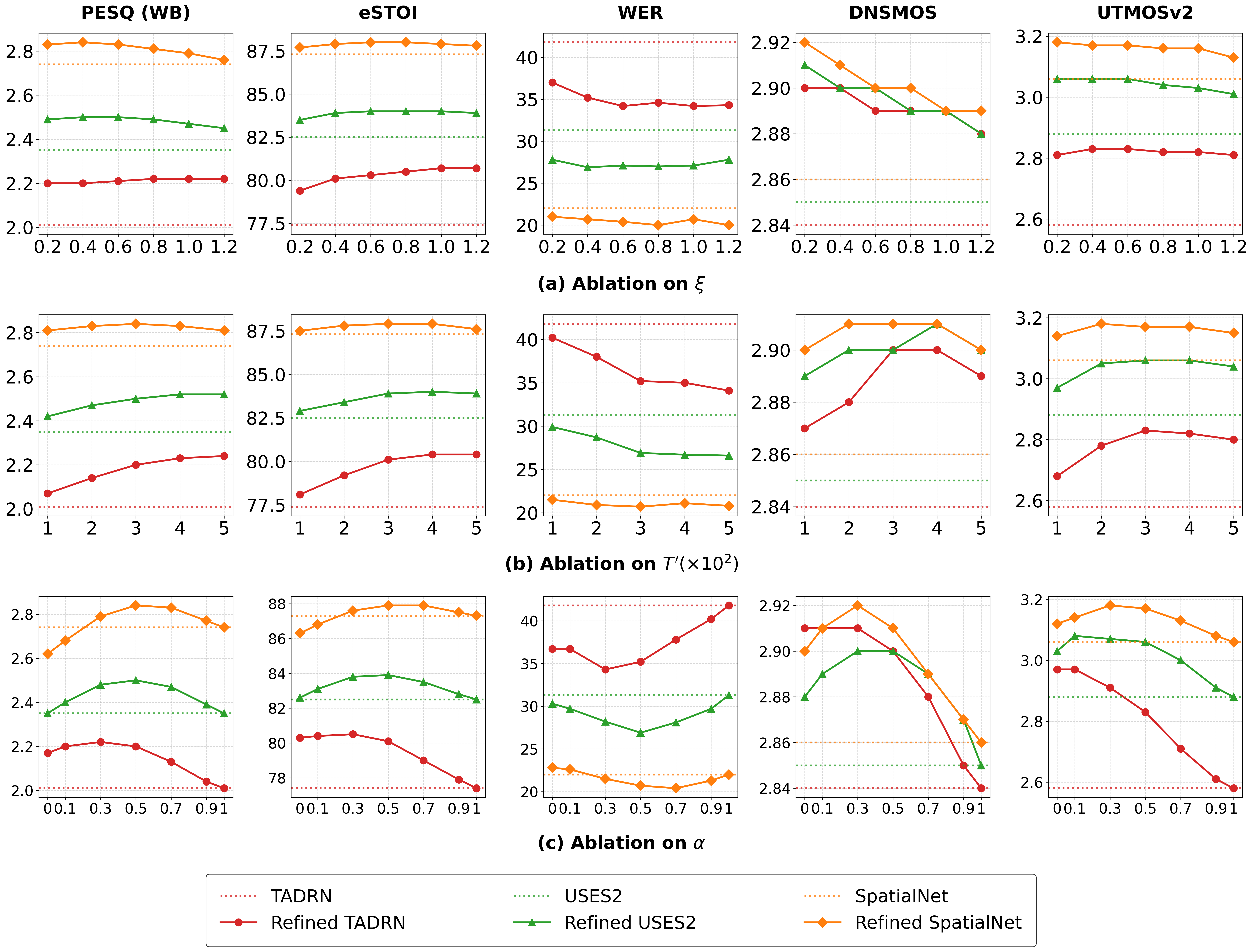}
    \label{fig:ablations}
\end{figure*}

\begin{table*}[t]
\scriptsize
\centering
\caption{Uni-ArrayDPS evaluation for multi-channel speech enhancement on the RealMan~\cite{realman} dataset.}
\setlength\tabcolsep{0.7pt}
\renewcommand{\arraystretch}{0.7}
\label{tab:enhancement_realman}
\begin{tabular}{@{}c c | c | c | ccccccc | ccccccc@{}}
\toprule
\multirowcell{2}{row} &
\multirowcell{2}{Methods} &
\multirowcell{2}{$\xi$} &
\multirowcell{2}{$\alpha$} &
\multicolumn{7}{c|}{\textbf{4-channel}} &
\multicolumn{7}{c}{\textbf{8-channel}} \\
\cmidrule(lr){5-11} \cmidrule(l){12-18}
& & & & STOI & eSTOI & { PESQ(NB/WB)} & SI-SDR & CER(\%) & DNSMOS & UTMOSv2 & STOI & eSTOI & {PESQ(NB/WB)} & SI-SDR & CER(\%) & {DNSMOS} & UTMOSv2 \\
\midrule
\rowcolor{gray!10}
A0 & Noisy & - & - & 0.712 & 0.533 & 1.58 / 1.11 & -6.1 & 56.6 & 1.72 & 2.09 & 0.712 & 0.533 & 1.58 / 1.11 & -6.1 & 56.6 & 1.72 & 2.09 \\
\midrule
A1 & TADRN \cite{tadrn} & - & - & 0.772 & 0.650 & 2.25 / 1.57 & -0.8 & 60.2 & 2.48 & 1.82 & 0.791 & 0.673 & 2.32 / 1.63 & -0.6 & 57.0 & 2.46 & 1.84 \\
\rowcolor{gray!10}
A2 & Refined TADRN & 0.4 & 0 & 0.770 & 0.646 & 2.32 / 1.63 & -1.0 & 59.4 & \textbf{2.63} & \textbf{2.28} & 0.793 & 0.673 & 2.40 / 1.70 & -0.8 & 55.4 & 2.60 & 2.29 \\
A3 & Refined TADRN & 0.4 & 0.5 & 0.781 & 0.664 & 2.35 / 1.67 & \textbf{-0.7} & 57.0 & 2.55 & 2.09 & \textbf{0.835} & \textbf{0.747} & \textbf{2.55} / \textbf{1.87} & \textbf{0.9} & 41.0 & 2.62 & \textbf{2.45} \\
\rowcolor{gray!10}
A4 & Refined TADRN & 0.2 & 0.5 & 0.776 & 0.656 & 2.32 / 1.65 & -0.8 & 58.9 & 2.56 & 2.07 & 0.832 & 0.743 & 2.54 / 1.86 & \textbf{0.9} & 42.2 & \textbf{2.63} & \textbf{2.45} \\
A5 & Refined TADRN & 0.6 & 0.5 & \textbf{0.784} & \textbf{0.669} & \textbf{2.37} / \textbf{1.68} & \textbf{-0.7} & \textbf{55.2} & 2.56 & 2.09 & 0.804 & 0.693 & 2.45 / 1.75 & -0.5 & \textbf{52.0} & 2.54 & 2.10 \\
\midrule
B1 & FaSNet-TAC \cite{fasnet} & - & - & 0.776 & 0.652 & 2.26 / 1.48 & -0.5 & 62.4 & 2.40 & 1.50 & 0.798 & 0.680 & 2.36 / 1.56 & 0.6 & 57.5 & 2.42 & 1.63 \\
\rowcolor{gray!10}
B2 & Refined FaSNet-TAC & 0.4 & 0 & 0.773 & 0.647 & 2.32 / 1.55 & -0.4 & 62.9 & \textbf{2.56} & \textbf{2.07} & 0.796 & 0.677 & 2.43 / 1.64 & 0.7 & 57.2 & \textbf{2.56} & \textbf{2.14} \\
B3 & Refined FaSNet-TAC & 0.4 & 0.5 & 0.788 & 0.668 & 2.37 / 1.58 & \textbf{-0.3} & 59.4 & 2.50 & 1.82 & 0.809 & 0.696 & \textbf{2.48} / \textbf{1.67} & \textbf{0.8} & \textbf{53.5} & 2.53 & 1.94 \\
\rowcolor{gray!10}
B4 & Refined FaSNet-TAC & 0.2 & 0.5 & 0.783 & 0.660 & 2.35 / 1.57 & -0.4 & 61.1 & 2.52 & 1.82 & 0.804 & 0.689 & 2.45 / 1.66 & 0.7 & 56.4 & 2.54 & 1.93 \\
B5 & Refined FaSNet-TAC & 0.6 & 0.5 & \textbf{0.791} & \textbf{0.673} & \textbf{2.39} / \textbf{1.59} & -0.2 & \textbf{58.4} & 2.51 & 1.83 & \textbf{0.811} & \textbf{0.704} & \textbf{2.48} / \textbf{1.67} & \textbf{0.8} & 53.9 & 2.53 & 1.93 \\
\midrule
C1 & USES2 \cite{uses2} & - & - & 0.863 & 0.778 & 2.80 / 2.09 & 1.09 & 40.2 & 2.69 & 2.71 & 0.876 & 0.798 & 2.87 / 2.19 & 2.2 & 35.6 & 2.71 & 2.78 \\
\rowcolor{gray!10}
C2 & Refined USES2 & 0.4 & 0 & 0.847 & 0.749 & 2.64 / 1.91 & 1.1 & 46.9 & 2.70 & 1.54 & 0.862 & 0.772 & 2.73 / 2.01 & 2.3 & 40.7 & 2.73 & 2.58 \\
C3 & Refined USES2 & 0.4 & 0.5 & 0.864 & 0.778 & \textbf{2.81} / \textbf{2.14} & \textbf{1.3} & 39.3 & 2.73 & \textbf{2.76} & 0.877 & 0.799 & \textbf{2.90} / 2.24 & \textbf{2.4} & 35.5 & 2.74 & 2.80 \\
\rowcolor{gray!10}
C4 & Refined USES2 & 0.2 & 0.5 & 0.862 & 0.775 & \textbf{2.81} / 2.13 & 1.2 & 40.8 & \textbf{2.74} & \textbf{2.76} & 0.875 & 0.795 & 2.88 / \textbf{2.24} & \textbf{2.4} & 35.7 & \textbf{2.76} & \textbf{2.81} \\
C5 & Refined USES2 & 0.6 & 0.5 & \textbf{0.865} & \textbf{0.780} & \textbf{2.81} / 2.12 & \textbf{1.3} & \textbf{39.0} & 2.71 & 2.75 & \textbf{0.878} & \textbf{0.800} & 2.89 / 2.23 & \textbf{2.4} & \textbf{33.7} & 2.73 & 2.79 \\
\midrule
D1 & SpatialNet \cite{spatialnet} & - & - & 0.853 & 0.770 & 2.62 / 1.74 & 2.2 & \textbf{40.1} & 2.64 & 2.48 & 0.831 & 0.743 & 2.47 / 1.71 & 0.8 & 41.5 & 2.51 & 2.36 \\
\rowcolor{gray!10}
D2 & Refined SpatialNet & 0.4 & 0 & 0.836 & 0.739 & 2.52 / 1.73 & 2.1 & 46.6 & 2.70 & 2.37 & 0.817 & 0.718 & 2.45 / 1.77 & 0.7 & 45.9 & 2.62 & 2.33 \\
D3 & Refined SpatialNet & 0.4 & 0.5 & 0.854 & 0.770 & \textbf{2.66} / \textbf{1.86} & \textbf{2.4} & 40.5 & 2.72 & 2.56 & 0.835 & 0.747 & \textbf{2.55} / \textbf{1.87} & \textbf{1.0} & 41.1 & 2.62 & \textbf{2.45} \\
\rowcolor{gray!10}
D4 & Refined SpatialNet & 0.2 & 0.5 & 0.853 & 0.767 & 2.65 / \textbf{1.86} & 2.3 & 41.5 & \textbf{2.73} & \textbf{2.57} & 0.832 & 0.743 & 2.54 / 1.86 & 0.9 & 42.5 & \textbf{2.63} & \textbf{2.45} \\
D5 & Refined SpatialNet & 0.6 & 0.5 & \textbf{0.856} & \textbf{0.772} & \textbf{2.66} / \textbf{1.86} & \textbf{2.4} & \textbf{40.1} & 2.71 & 2.55 & \textbf{0.837} & \textbf{0.749} & \textbf{2.55} / 1.86 & \textbf{1.0} & \textbf{40.5} & 2.61 & \textbf{2.45} \\
\bottomrule
\end{tabular}
\end{table*}

\subsection{Enhancement Results and Analysis}
In this section, we first show the enhancement results on the simulated enhancement dataset described in Sec.~\ref{sec:enhancement_simulated}. Then we discuss the effects of the likelihood guidance $\xi$, the discriminative-generative interpolation coefficient $\alpha$, and the starting diffusion time $T'$, in Sec.~\ref{sec:ablations}. In the end, we discuss the enhancement results on the RealMan dataset in Sec.~\ref{sec:enhancement_realman}.
\subsubsection{Simulated Dataset}\label{sec:enhancement_simulated}
We first show the multi-channel enhancement evaluation results on our simulated test dataset, in Table~\ref{tab:enhancement_simulated}. By observing the metrics of the baseline discriminative models and the Uni-ArrayDPS refinement, it is clear that our refinement method (row A3, B3, C3, D3 are default configurations) is able to consistently improve the corresponding discriminative model, in all evaluation metrics.

By observing TADRN's results in row A3, Uni-ArrayDPS refined TADRN can improve the original TADRN by about 0.03 in eSTOI, 0.2 in wide-band PESQ, 1 dB in SI-SDR, 0.2 in UTMOSv2, and 6 percent in WER. In row A2, $\alpha=0$ means that the diffusion posterior sampling result is directly used and there is no discriminative-generative interpolation, which shows much higher UTMOSv2 score. The same also happens to refined FaSNet-TAC in row B2, which is probably because time-domain architectures suffer in perceptual quality. row A5 further shows that higher likelihood guidance $\xi$ can improve speech intelligibility, since higher $\xi$ means using more mixture information. We will further discuss the effects of the likelihood $\xi$, starting diffusion steps $T'$ in Sec.~\ref{sec:ablations}, Fig.~\ref{fig:ablations}.

For FaSNet-TAC's result in row B1-B5, we can see that Uni-ArrayDPS's default setting in row B3, can consistently improve over the original FaSNet-TAC. It can improve the baseline by about 0.04 in eSTOI, 0.2 in wide-band PESQ, 10 per cent in WER, and 0.5 in UTMOSv2. row B2 ($\alpha=0$) shows even better improvements.

For USES2's result in row C1-C5, the default Uni-ArrayDPS result in row C3 can improve original USES2 by about 0.015 in eSTOI, 0.15 in wide-band PESQ, 4 percent in WER, and 0.1 in UTMOSv2, for both 4-channel and 8-channel cases.

Similarly, for SpatialNet's result in row D1-D5, Uni-ArrayDPS refinement can still improve the strong SpatialNet-Large model consistently in all metrics. As shown in row D3, Uni-ArrayDPS improves SpatialNet by about 0.01 in eSTOI, 0.1 in wide-band PESQ, 1 percent in WER, and 0.1 in UTMOSv2.

\subsubsection{Ablation Studies}\label{sec:ablations}
As shown in Table~\ref{tab:enhancement_simulated}, different configurations of the discriminative-generative interpolation coefficient $\alpha$, likelihood guidance $\xi$ can have an influence on the refinement performance. Also, note that as mentioned in Sec.~\ref{sec:config}, we set $T'=300$ by default, which is also applied for all the experiments in Table~\ref{tab:enhancement_simulated}. Thus, we study the effects of these parameters in Fig.~\ref{fig:ablations}, where row (a) in Fig.~\ref{fig:ablations} shows the ablations on $\xi\in\{02, 0.4, 0.6, 0.8, 1.0, 1.2\}$, row (b) shows the ablations on $T'\in\{100, 200, 300, 400, 500\}$, and row (c) shows the ablations on $\alpha\in\{0, 0.1, 0.3, 0.5, 0.7, 0.9, 1.0\}$. All results in Fig.~\ref{fig:ablations} are from the 4-channel enhancement experiments on the simulated dataset.

In row (a) of Fig.~\ref{fig:ablations}, we can observe five different metrics' relation with the likelihood guidance $\xi$. Note that $\xi$ is a guidance term in line 16 of Algorithm~\ref{alg:inference}, which determines how much likelihood guidance is used in each posterior sampling step. Intuitively, a higher $\xi$ would result in enhanced outputs more complied with the original mixture, minimizing hallucination effects caused by the diffusion generation. This can then be confirmed in row (a) of Fig.~\ref{fig:ablations}. We can see that a higher $\xi$ have higher eSTOI and lower WER than lower $\xi$s, meaning that increasing $\xi$ tends to improve speech intelligibility. This phenomenon is very obvious for TADRN and USES2, but more subtle for SpatialNet. On the other hand, high $\xi$ might include more noise from the noisy mixtures, resulting in noisier results. This can be verified in the PESQ, DNSMOS, and UTMOS chart in row (a) of Fig.~\ref{fig:ablations}, where these metrics degrade as $\xi$ increase. This pattern is extremely obvious for SpatialNet and USES2, and less obvious for TADRN. Overall, the likelihood guidance $\xi$ can be used as a knob to balance the tradeoff between speech intelligibility and perceptual quality, towards solving the well-known hallucination problem in generative speech enhancement.

Row (b) of Fig.~\ref{fig:ablations} shows the different metrics with respect to the starting diffusion step $T'$ introduced in Uni-ArrayDPS algorithm. $T'$ determines at what diffusion time for Uni-ArrayDPS to start. For the two extreme cases, if $T'=0$, then Uni-ArrayDPS does not refine anything and just return the discriminative model's output. If $T'=T=1000$, then we start from Gaussian noise and our likelihood approximation in Sec.~\ref{sec:uniarraydps_derivation} would not be accurate, causing convergence issue~\cite{arraydps}. If we start from our default $T'=300$, then we start our diffusion process from an initialization, which is a weighted sum of the discriminative model's output and Gaussian noise, and then Uni-ArrayDPS will learn to recover the information masked from the Gaussian noise, using the speech prior information along with the multi-channel mixtures. Thus, a higher $T'$ means more noise in the initialization and more room to process. In Fig.~\ref{fig:ablations} row (b), we can see that PESQ, eSTOI, and WER improve when $T'$ increases from $100$ to $300$, and then roughly stay flat. DNSMOS and UTMOSv2 then increase or stay flat as $T'$ increases to $400$, and then starts to degrade when $T'=500$. These finding shows that it is best to start from $T'=300$ or $400$ steps, which not only provides enough room for refinement, but also prevents refinement from a very noisy initialization.

Lastly, Fig.~\ref{fig:ablations} row (c) shows ablations on the discriminative-generative interpolation coefficient $\alpha$, which is introduced in Algorithm~\ref{alg:inference} (line 23). The coefficient $\alpha$ interpolates between the diffusion posterior sampling (DPS) output and the discriminative model's output. Thus, $\alpha=1$ corresponds to using only the discriminative model's output, whereas $\alpha=0$ corresponds to using only the DPS output. From Fig.~\ref{fig:ablations} row (c), we observe that for all models, as $\alpha$ increases, the metrics first improve and then start to degrade. For TADRN and USES2, most interpolation coefficients yield consistent improvements on most metrics. However, for SpatialNet, Uni-ArrayDPS improves SpatialNet in PESQ, eSTOI, and WER only when $\alpha>0.3$. Interestingly, UTMOS and DNSMOS tend to be better when $\alpha$ is small, suggesting that the DPS output has higher perceptual quality than the discriminative models. Overall, $\alpha=0.5$ is a safe choice that enables Uni-ArrayDPS to consistently improve different models across metrics.

\subsubsection{RealMan Dataset}\label{sec:enhancement_realman}
This section shows the multi-channel enhancement evaluation results on the RealMan test dataset discussed in Sec.~\ref{sec:datasets_metrics}, in Table~\ref{tab:enhancement_realman}. Similar to the enhancement result in the simulated dataset in Table~\ref{tab:enhancement_simulated}, Uni-ArrayDPS can also consistently improve over any discriminative models, in all metrics. Note that the prior diffusion model is trained on DNS-Challenge, which is in English, while the RealMan dataset is in Chinese. This further shows Uni-ArrayDPS's domain generalization abilities. From Table~\ref{tab:enhancement_realman}, we can observe that Uni-ArrayDPS provides consistent gains on this real-world recorded dataset. Note that for RealMan (Chinese), the ASR metric is character error rate (CER) instead of word error rate (WER).

For TADRN's results in row A1-A5, Uni-ArrayDPS consistently improves TADRN in both 4-channel and 8-channel settings. In particular, for the 8-channel case, the default configuration (row A3) improves the TADRN (row A1) by 0.07 in eSTOI, 0.24 in wide-band PESQ, 16 percent in CER, and 0.61 in UTMOSv2. The improvement is less pronounced for the 4-channel case, but is still consistent.

For FaSNet-TAC's results in row B1-B5, we can see that the default configuration (row B3) improves FaSNet-TAC by about 0.02 in eSTOI, 0.01 in Wide-Band PESQ, 4 percent in CER, and 0.3 in UTMOSv2. Similar improvements are also shown for 4-channel enhancement. One interesting observation is that for the simulated dataset's result in Table~\ref{tab:enhancement_simulated}, TADRN outperforms FaSNet-TAC by a large margin, while here in Table~\ref{tab:enhancement_realman}, FaSNet-TAC has much better performance when generalizing to the RealMan dataset.

For USES2's results in row C1-C5, USES (row C1) also shows great generalization ability towards the RealMan dataset, performing even better than the SpatialNet (row D1). Uni-ArrayDPS's default refinement (row C3) improves USES2 by 0.016 in eSTOI, about 0.1 in wide-band PESQ, 1 percent in CER, and 0.05 in UTMOS v2 for the 4-channel case. Similar results are also shown for the 8-channel case. Also, with a larger $\xi$, row C5 shows much better improvement in CER.

Similarly, for SpatialNet's results in row D1-D5, default Uni-ArrayDPS (row D3) improves SpatialNet (D1) by about 0.1 in wide-band PESQ, and 0.1 for UTMOSv2. The improvement is mainly for perceptual quality and very subtle for intelligibility.


\begin{table*}[t]
\scriptsize
\centering
\caption{Uni-ArrayDPS evaluation for multi-channel speech separation and enhancement on the simulated adhoc microphone array dataset.}
\setlength\tabcolsep{0.7pt}
\renewcommand{\arraystretch}{0.7}
\label{tab:separation_simulated}
\begin{tabular}{@{}c c | c | c | ccccccc | ccccccc@{}}
\toprule
\multirowcell{2}{row} &
\multirowcell{2}{Methods} &
\multirowcell{2}{$\xi$} &
\multirowcell{2}{$\alpha$} &
\multicolumn{7}{c|}{\textbf{4-channel}} &
\multicolumn{7}{c}{\textbf{8-channel}} \\
\cmidrule(lr){5-11} \cmidrule(l){12-18}
& & & & STOI & eSTOI & { PESQ(NB/WB)} & SI-SDR & WER(\%) & DNSMOS & UTMOSv2 & STOI & eSTOI & {PESQ(NB/WB)} & SI-SDR & WER(\%) & {DNSMOS} & UTMOSv2 \\
\midrule
\rowcolor{gray!10}
A0 & Noisy & - & - & 0.568 & 0.299 & 1.29 / 1.09 & -10.1 & 95.8 & 1.47 & 1.66 & 0.568 & 0.299 & 1.29 / 1.09 & -10.1 & 95.8 & 1.47 & 1.66 \\
\midrule
B1 & FaSNet-TAC \cite{fasnet} & - & - & 0.755 & 0.540 & 2.06 / 1.35 & 1.63 & 75.0 & 2.19 & 1.26 & 0.780 & 0.577 & 2.15 / 1.40 & 2.60 & 69.1 & 2.29 & 1.35 \\
\rowcolor{gray!10}
B2 & Refined FaSNet-TAC & 0.4 & 0 & \textbf{0.786} & \textbf{0.607} & \textbf{2.31} / \textbf{1.52} & \textbf{2.71} & \textbf{63.1} & \textbf{2.54} & \textbf{2.10} & \textbf{0.820} & \textbf{0.657} & \textbf{2.45} / \textbf{1.63} & \textbf{3.90} & \textbf{53.9} & \textbf{2.59} & \textbf{2.23} \\
B3 & Refined FaSNet-TAC & 0.4 & 0.5 & 0.781 & 0.587 & 2.27 / 1.48 & 2.37 & 63.4 & 2.39 & 2.68 & 0.808 & 0.630 & 2.39 / 1.57 & 3.40 & 55.4 & 2.43 & 1.81 \\
\rowcolor{gray!10}
B4 & Refined FaSNet-TAC & 0.2 & 0.5 & 0.775 & 0.576 & 2.24 / 1.47 & 2.21 & 66.5 & 2.39 & 1.67 & 0.801 & 0.618 & 2.39 / 1.57 & 3.40 & 58.5 & 2.44 & 1.80 \\
B5 & Refined FaSNet-TAC & 0.6 & 0.5 & 0.784 & 0.594 & 2.28 / 1.49 & 2.43 & 62.3 & 2.38 & 1.68 & 0.812 & 0.637 & 2.40 / 1.59 & 3.50 & \textbf{53.7} & 2.43 & 1.79 \\
\midrule
C1 & USES2 \cite{uses2} & - & - & 0.880 & 0.754 & 2.81 / 2.00 & 4.25 & 42.5 & 2.75 & 2.50 & 0.892 & 0.775 & 2.91 / 2.09 & 4.20 & 37.7 & 2.75 & 2.62 \\
\rowcolor{gray!10}
C2 & Refined USES2 & 0.4 & 0 & 0.889 & 0.776 & 2.88 / 2.10 & \textbf{5.10} & 37.9 & \textbf{2.90} & \textbf{2.90} & 0.907 & 0.806 & 3.00 / 2.25 & \textbf{4.90} & 32.4 & \textbf{2.90} & \textbf{2.96} \\
C3 & Refined USES2 & 0.4 & 0.5 & \textbf{0.894} & 0.779 & \textbf{2.96} / \textbf{2.19} & 4.92 & 35.9 & 2.85 & 2.80 & 0.908 & 0.805 & 3.06 / 2.32 & 4.70 & 31.1 & 2.86 & 2.89 \\
\rowcolor{gray!10}
C4 & Refined USES2 & 0.2 & 0.5 & 0.890 & 0.774 & 2.94 / 2.17 & 4.80 & 38.1 & 2.85 & 2.79 & 0.905 & 0.800 & 3.05 / 2.30 & 4.70 & 32.0 & 2.86 & 2.88 \\
C5 & Refined USES2 & 0.6 & 0.5 & \textbf{0.895} & \textbf{0.782} & \textbf{2.96} / \textbf{2.20} & 4.90 & \textbf{36.0} & 2.84 & 2.79 & \textbf{0.909} & \textbf{0.808} & \textbf{3.07} / \textbf{2.33} & 4.70 & \textbf{30.4} & 2.85 & 2.88 \\
\midrule
D1 & SpatialNet \cite{spatialnet} & - & - & 0.933 & 0.848 & 3.19 / 2.53 & 12.20 & 26.3 & 2.87 & 2.97 & 0.952 & 0.887 & 3.39 / 2.80 & 13.80 & 19.8 & 2.89 & 3.09 \\
\rowcolor{gray!10}
D2 & Refined SpatialNet & 0.4 & 0 & 0.929 & 0.846 & 3.15 / 2.49 & 12.10 & 26.2 & 2.93 & 3.06 & 0.947 & 0.879 & 3.32 / 2.74 & 13.50 & 20.4 & 2.93 & 3.10 \\
D3 & Refined SpatialNet & 0.4 & 0.5 & \textbf{0.938} & \textbf{0.861} & \textbf{3.28} / \textbf{2.70} & 12.70 & 23.1 & 2.94 & \textbf{3.11} & 0.955 & 0.894 & \textbf{3.48} / \textbf{2.98} & 14.30 & 17.8 & 2.95 & \textbf{3.18} \\
\rowcolor{gray!10}
D4 & Refined SpatialNet & 0.2 & 0.5 & 0.937 & 0.858 & \textbf{3.28} / \textbf{2.70} & 12.60 & 23.8 & \textbf{2.95} & \textbf{3.11} & 0.954 & 0.892 & 3.47 / 2.97 & 14.10 & 18.2 & \textbf{2.96} & \textbf{3.18} \\
D5 & Refined SpatialNet & 0.6 & 0.5 & \textbf{0.938} & \textbf{0.861} & \textbf{3.28} / 2.69 & \textbf{12.80} & \textbf{23.3} & 2.93 & 3.10 & \textbf{0.956} & \textbf{0.896} & \textbf{3.48} / \textbf{2.98} & \textbf{14.40} & \textbf{17.4} & 2.94 & 3.17 \\
\bottomrule
\end{tabular}
\end{table*}
\subsection{Separation Results and Analysis}
We show the results for multi-channel 2-speech separation in Table~\ref{tab:separation_simulated}, with the simulated noisy separation dataset mentioned in Sec.~\ref{sec:baseline}.

From row B1-B5 in Table~\ref{tab:separation_simulated}, we can see that $\alpha=0$ (row B2) shows the best refinement performance, which improves the FaSNet-TAC baseline by about 0.08 in eSTOI, 0.23 in wide-band PESQ, 0.3 in narrow-band PESQ, 1.3 dB in SI-SDR, more than 15 percent in WER, and about 0.9 in UTMOSv2, for the 8-channel case. The default configuration in row B3 also shows remarkable results for all metrics, showing Uni-ArrayDPS's ability to adapt to source separation in noisy environments.

For USES2's separation results from row C1 to C5, the default Uni-ArrayDPS (row C3) improves the USES2 baseline by 0.03 in eSTOI, more than 0.2 in wide-band PESQ, more than 6 percent in WER, and about 0.3 in UTMOSv2 for the 8-channel case. Similar improvement is also shown for the 4-channel case and other parameter configurations.

For the strongest baseline in simulated datasets, SpatialNet can also be greatly improved by Uni-ArrayDPS for noisy source separation. In row D3, the default Uni-ArrayDPS can improve SpatialNet by more than 0.01 in eSTOI, about 0.2 in wide-band PESQ, more than 3 percent in WER, and 0.1 in UTMOSv2 for the 4-channel setting.

Overall, extensive results have shown that Uni-ArrayDPS can refine any strong and competitive discriminative model, for both multi-channel speech enhancement and separation. The improvement can be observed for both intelligibility and perceptual metrics. We also show ablations on different parameters and how they would affect the algorithm.

\section{Conclusion}

We introduced Uni-ArrayDPS, a training-free, generative, and array-agnostic refinement framework that leverages a pre-trained speech diffusion model to improve the outputs of existing discriminative multi-channel enhancement and separation systems. Starting from a discriminative estimate, we estimate a noise spatial covariance matrix and use that to guide an ArrayDPS sampling procedure that enforces multi-channel consistency while steering the generative prior toward clean speech.

Across a range of backbones (including SOTA time-domain and STFT-domain baselines) and array configurations (e.g., 4- and 8-channel setups), ArrayDPS-Refine consistently improves both perceptual quality and intelligibility, showing convincing results in both intrusive metrics (SI-SDNR, PESQ, STOI, WER), and non-instrusive metrics (DNSMOS, UTMOSv2). These results indicate that Uni-ArrayDPS refinement can serve as a practical plug-and-play module for multi-microphone speech processing without any constraint of the discriminative model, array-geometry, and number of sources.

However, there are a few limitations of Uni-ArrayDPS. First, Uni-ArrayDPS is based on diffusion posterior sampling, which is computationally expensive and unsuitable for real-time processing. Second, we assume static speakers in this paper, while moving sources are common in real-world scenarios. We leave these limitations to future research.


%





\ifCLASSOPTIONcaptionsoff
  \newpage
\fi



\bibliographystyle{IEEEtran}
\bibliography{refs}
\end{document}